\documentclass[12pt]{article}

\usepackage[hyphens]{url}
\usepackage[utf8]{inputenc}
\usepackage{charter}
\usepackage{fullpage}
\usepackage{graphicx}
\usepackage{lipsum}
\usepackage[sort&compress,numbers]{natbib}
\usepackage[colorlinks=true,linkcolor=blue,citecolor=blue,urlcolor=blue]{hyperref}
\usepackage[total={6.5in,9in},top=1in,headsep=0.1in,headheight=1in]{geometry}
\usepackage{enumitem}
\usepackage{amssymb}
\usepackage{subfigure}

\newcommand\snowmass{
\begin{center}
  \rule[-0.2in]{\hsize}{0.01in}\\
  \rule{\hsize}{0.01in}\\
  \vskip 0.1in
  Submitted to the Proceedings of the US Community Study\\ 
  on the Future of Particle Physics (Snowmass 2021)\\
  \rule{\hsize}{0.01in}\\
  \rule[+0.2in]{\hsize}{0.01in}\\[-2em]
\end{center}
}

\usepackage[firstpage=true]{background}
\backgroundsetup{contents={\parbox{6.5in}{\snowmass}}, scale=1,placement=top,opacity=1,color=black,position={3.25in,1.2in}}

\usepackage{fancyhdr}
\fancypagestyle{plain}{%
  \fancyhf{}%
  \fancyhead[C]{}
  \fancyfoot[C]{\thepage}
}

\fancypagestyle{empty}{%
  \fancyhf{}%
  \fancyhead[C]{{\it Snowmass 2021 CEF06: Public Policy \& Government Engagement}}
  \fancyfoot[C]{\thepage}
}
\usepackage{authblk}

\usepackage{comment}

\title{Snowmass '21 Community Engagement Frontier 6: Public Policy and Government Engagement\\ Congressional Advocacy for Areas Beyond HEP Funding
}

\date{}

\author[1]{Richie Diurba}
\author[2]{Rob Fine}
\author[3]{Mandeep Gill}
\author[4]{Harvey Newman}
\author[5]{Kevin Pedro}
\author[6]{Alexx Perloff}
\author[7]{Breese Quinn}
\author[5]{Louise Suter}
\author[8]{Shawn Westerdale}

\affil[1]{Universit\"{a}t Bern}
\affil[2]{Los Alamos National Laboratory}
\affil[3]{Kavli Institute}
\affil[4]{California Institute of Technology}
\affil[5]{Fermi National Accelerator Laboratory}
\affil[6]{University of Colorado Boulder}
\affil[7]{University of Mississippi}
\affil[8]{Princeton University}

\begin{document}

\maketitle
\tableofcontents

\section{Introduction}
\label{sec:intro}

This document has been prepared as a Snowmass contributed paper by the Public Policy \& Government Engagement topical group (CEF06) within the Community Engagement Frontier. The charge of CEF06 is to review all aspects of how the High Energy Physics (HEP) community engages with government at all levels and how public policy impacts members of the community and the community at large, and to assess and raise awareness within the community of direct community-driven engagement of the US federal government (\textit{i.e.} advocacy). In the previous Snowmass process these topics were included in a broader ``Community Engagement and Outreach'' group whose work culminated in the recommendations outlined in Ref. \cite{snowmass13recs}.

The focus of this paper is the potential for HEP community advocacy on topics other than funding for basic research. The HEP community has run a very active and successful advocacy effort for many years that is targeted at the U.S. federal legislature (\textit{i.e.} Congress) and is focused on funding for HEP and basic research overall. The full details of that effort, as well as ideas for expanding  community advocacy for funding to other government entities, are described in two other contributed papers prepared by CEF06 \cite{cef06paper1,cef06paper3}. Throughout this Snowmass process, one of the main themes in the activities of this group was focused on policy issues that directly impact the HEP community at large, as well as individual members of the community, which are not directly related to the funding for our field. In our discussions on this topic, we attempted to address the following questions: What should be our role as a community, and individuals, in trying to influence legislation about these issues? What advocacy may we already be engaged in as a community, and as individuals, pertaining to these issues? What resources already exist to advocate for these issues? And, should the HEP community organize an explicit effort to advocate for these issues? We note that no consensus was ultimately reached on answers to some of these questions. We therefore do not present answers, but rather a summary of the discussions that took place in the framework we developed to discuss this matter most thoroughly.

The paper is organized as follows. Section \ref{sec:important} describes legislative topics that the community already implicitly advocates for. Section \ref{sec:resources} describes currently available resources for non-funding advocacy and how they could be better utilized by members of the community. Section \ref{sec:discussion} summarizes the discussions that took place on this topic as part of the Snowmass process. 
\section{Legislative Topics that Impact Particle Physics}
\label{sec:important}

Many topics beyond federal research funding significantly impact HEP research and researchers. We provide a brief overview of these topics here. This section does not include an exhaustive list of issues or impacts. Additional details are available in the cited references where available. 

``Diversity, equity, and inclusion'' (DEI) has become a catch-all term for a broad array of issues related to improving HEP physicists' access to research, experience performing research, and success in the field. In particular, expanding access to resources to perform research has been identified as crucial for the health of the field. Many small colleges and universities have historically had limited, if any, opportunities to participate in research. This group includes historically Black colleges and universities (HBCUs), other minority-serving institutions (MSIs), and other small institutions, such as liberal-arts colleges and others without graduate programs. The federal government's various EPSCoR (Established Program to Stimulate Competitive Research) programs take a step towards addressing these inequalities, yet their focus on state-level action neglects small colleges and universities in states with large institutions that receive substantial funding. The new America Creating Opportunities for Manufacturing, Pre-Eminence in Technology, and Economic Strength (COMPETES) Act, recently passed by the U.S. House of Representatives, includes a promising program to increase funding to small institutions more broadly, without restrictions based on state~\cite{Ambrose2022feb}.
Further discussion of DEI issues can be found in the proceedings of CEF03, the Snowmass topical group for Diversity \& Inclusion.~\cite{CEF03}.

% any statistics on how many US HEP researchers are immigrants?
The international nature of HEP makes immigration a vital issue. Many scientists from other countries come to the U.S. to pursue Ph.D., postdoctoral, or faculty positions or to collaborate with U.S. scientists and use U.S. research facilities. The uncertainty and opacity inherent in the current U.S. immigration system imposes substantial physical and psychological costs on HEP researchers and reduces the field's efficiency and productivity. The widespread practice of educating Ph.D. researchers in the U.S. and refusing to extend their visas or grant them permanent residency harms national competitiveness in R\&D. Attracting and retaining talented researchers, regardless of their national origin, should be considered a matter of importance for national security. The aforementioned America COMPETES Act includes, as an amendment, an approach that would be revolutionary for HEP and U.S. scientific research as a whole. This approach is the Keep STEM Talent Act~\cite{Foster2019}, which makes immigrants who earned Masters or Ph.D. degrees in STEM fields automatically eligible to apply for permanent residency in the U.S. This simple change, which has been proposed repeatedly in the past, would substantially improve the prospects and experiences for foreign researchers who seek to learn and work in the U.S. Other practical improvements to the immigration process are also desirable, such as waiving repeated interviews and costly, time-consuming visa stamping procedures that require recipients to leave and reenter the U.S.

Related issues of so-called ``research security'' have experienced increasing emphasis recently, with largely detrimental consequences. While the need for extreme care and caution is widely acknowledged for classified topics, basic research has benefited from openness and collaboration for many decades, including throughout the Cold War. The recent introduction of vaguely-defined ``sensitive'' areas of research and corresponding concerns about ``foreign talent recruitment programs'' have had a widely-reported chilling effect on international collaboration, especially with Chinese institutions and researchers~\cite{APS2021,Thomas2022jan}. Related legal actions have primarily focused on Chinese citizens who live and work as faculty and researchers in the U.S., rising to the level of persecution~\cite{C100wp,amicus,Thomas2021jul}. Punishing researchers for pursuing their research harms U.S. national security, rather than helping it, by driving away talented scientists. Instead, the government should provide researchers with adequate funding to eliminate the incentive to seek support from possibly malign foreign programs. A more comprehensive argument for increasing research funding is covered  in Ref.~\cite{cef06paper1}, while the argument for broadening access to funding is discussed above. To make efficient and legal use of federal research funding, it is still essential to avoid conflicts of interest and to avoid obtaining multiple, overlapping sources of support for the same project, sometimes called ``double-dipping''. Therefore, funding disclosure rules should be clarified and procedures streamlined, with any criminal implications removed.

There are other, longer-standing issues regarding immigration status and HEP research in the U.S. National laboratories drive a substantial amount of HEP experimental research, and even those nominally open to the public often have stricter access restrictions imposed than other institutions (such as universities). Individuals who originate from ``countries of concern'' may be barred from accessing otherwise publicly available resources based on a background check for which there is no transparency and no appeal.\footnote{We note a lack of readily available public documentation of this process, which in itself is reflective of poor transparency. Rather, we cite anecdotal evidence gleaned in interactions between some authors of this paper and members of the community.} Researchers visiting for workshops or conferences may be denied visas without transparency or appeal. Denying these visas does not just harm those researchers, but also U.S. HEP as a whole: critical research may be delayed or even abandoned, and the international community increasingly views the U.S. as an inhospitable location for important meetings of the field. Ultimately, national security will be best promoted by ensuring that procedures do not reject talented scientists. The past few decades of increasing international collaboration have been highly successful for HEP, and this approach should be expanded rather than discouraged.

\subsection{Implicit Advocacy in Community Materials}

Many of the above issues are touched on in the materials that the community uses in its current advocacy efforts. The main focus of the existing HEP advocacy efforts is to increase support for and knowledge about basic research, and HEP research specifically. While materials that touch on the above issues are used as part of the more extensive discussion about the benefits of supporting HEP, the community effort does not include advocacy for any specific policies or legislation in these areas. Similarly to the community advocacy materials focused on funding, the materials and the message about non-funding topics are crafted by a small group of designees of the the Fermilab Users Executive Committee (UEC), SLAC Users Organization (SLUO), and the U.S. Large Hadron Collider Users Association (USLUA), with input from APS Division of Particles and Fields (APS DPF). The membership of this group is coordinated by, but is not necessarily itself comprised of, elected representatives of our field (\textit{i.e.} the leadership of the users groups). It has been noted (see Ref. \cite{cef06paper1}) that this group is not fully representative of our field, and it does not ask for broader community input when crafting these messages. These topics all fall within a category of issues that have general community support, but the question remains if more explicit consensus-building is needed or may be appropriate.

The follow having been identified as areas in which the community implicitly advocates as a core component of existing advocacy efforts (this is not an exhaustive list):
\begin{itemize}
    \item STEM education
    \item International collaboration and international and open science 
    \item DEI impacts and efforts 
    \item Emerging technologies, QIS and AI
    \item Basic research support as a driver of U.S. innovation, economic development, and national security
\end{itemize}

Below is a list of community-produced materials that are used for existing advocacy efforts and which touch on the above non-funding issues. These materials can be found on the U.S. Particle Physics website \cite{uspp}, and are jointly maintained by APS DPF, UEC, SLUO, and USLUA.

\begin{itemize}
    \item \textbf{Particle Physics Makes a Difference in Your Life}
    \begin{itemize}
        \item Relevant issue: impact of basic research.
        \item Summary of content: the discoveries and technological developments made by the HEP community have an impact on the everyday lives of ordinary citizens. We know that this type of broad applicability resonates with members of Congress and their staff. This document discusses how HEP has impacted other fields and industries.
    \end{itemize}
    \item \textbf{Particle Physics Builds STEM Leaders}
    \begin{itemize}
        \item Relevant issue: STEM outreach.
        \item Summary of content: this document discusses the outreach and public engagement activities of the field. While education is not the top issue for voters \cite{pew2022jun,gallup2022mar}, it is a consistently important topic among their representatives and the funding agencies.
    \end{itemize}
    \item \textbf{Particle Physicists Value Diversity and Strive Toward Equity}
    \begin{itemize}
        \item Relevant issue: DEI outreach.
        \item Summary of content: most importantly, this document contains a statement of our community values and what the field is attempting to achieve in terms of diversity, equity, and inclusion. This document includes a clear statement that the community has work to do to improve the climate and opportunities for historically underrepresented groups (HUGs), and highlights a few programs explicitly designed to empower and provide opportunities for HUG members.
    \end{itemize}
    \item \textbf{Particle Physicists Deliver Discovery Science Through Collaboration}
    \begin{itemize}
        \item Relevant issues: international science; big science.
        \item Summary of content: this document discusses the various institution types and locations where HEP researchers work. The goal is to show that HEP is an international effort and involves the best minds from across the country and the world. This document also contains a summary of the projects outlined in the 2013 P5 plan, their timelines, and whether each is on schedule.
    \end{itemize}
    \item \textbf{Particle Physicists Advance Artificial Intelligence}
    \begin{itemize}
        \item Relevant issue: emerging technologies.
        \item Summary of content: although the HEP community has been working on machine learning (ML) for decades, artificial intelligence (AI) only became a national initiative in March 2020 \cite{naiia2020,aigov2022}. This document was created to explain how the HEP community uses ML and successfully interfaces with industrial partners to push the boundaries of ML research and development.
    \end{itemize}
    \item \textbf{Particle Physics and Quantum Information Science}
    \begin{itemize}
        \item Relevant issue: emerging technologies.
        \item Summary of content: Quantum information science (QIS) is another area of national interest \cite{osti_qis_2022mar,qi2022mar}. This document discusses the benefits of QIS, the skill sets that make HEP researchers valuable in these endeavors, and how QIS developments can in turn help solve fundamental problems in HEP research.
    \end{itemize}
\end{itemize}
    
\section{Government Relations Resources Available to the Community}
\label{sec:resources}

The HEP community has several ways in which to advocate for funding; this is discussed in detail in Ref.~\cite{cef06paper1}. As described in Section \ref{sec:important}, community resources that are used in this funding advocacy also implicitly advocate for other topics. Additional resources exist within the broader physics community to more explicitly advocate for the topics covered in this paper and others, and these resources are available to members of the HEP community. In many ways the groups which represent the larger physics community have considerably greater resources than the HEP community alone. However, the priorities of the broader physics community are not always aligned with those of the HEP community, and in practice there may be limitations to the use of these additional resources that need to be considered.

The American Physical Society (APS) is a nonprofit membership organization representing physicists within the U.S. The mission of this group is to advance and disseminate physics knowledge and advocate for the needs of physicists and scientists at large. More relevant to this paper, APS advocates for many non-funding legislative issues of importance to the physics community. However, the HEP community is only a subset of the broader APS membership, meaning that the influence of the HEP community over the issues taken up by APS is somewhat limited. Additionally APS, as a rule, will not elevate the interests of one division (APS DPF, in the case of HEP) over another in its advocacy.

There are a number of specific ways in which the HEP community can strengthen its connection to APS advocacy efforts. APS is a trade organization available to all physicists, though many members of the HEP community do not understand how the APS government affairs team makes their decisions or the infrastructure used to elect representatives. APS experts have been responsive in the past, so this may be due to a failure of the HEP community to highlight their work. \textbf{The HEP community should invest effort into building stronger connections with the APS team.} APS also offers Congressional Science Fellowships, which aid Congress by providing scientifically literate, skilled personnel. In return, the fellows have the opportunity to enhance their careers and to assist the physics community in better communicating with members of Congress. \textbf{The HEP community should raise awareness of this program and strive to increase the number of HEP community members within the federal government.}

The American Institute of Physics (AIP) is another organization that works to promote and advance the physical sciences. AIP is an umbrella organization that pools together the resources of ten member societies, including APS. As an umbrella organization, the AIP can help coordinate its individual members' activities and messages. We note that AIP has its own Congressional fellowship program. \textbf{The activities of AIP should be promoted within the HEP community.}

The American Association for the Advancement of Science (AAAS) is best known through its journal \textit{Science} and its government fellowship program. Its goal of the advancement of science means that it has a very active government engagement group with its own training programs and significant resources. AAAS hosts various efforts to get scientists involved in policy, such as the Local Science Engagement Network and National Science Policy Network. Perhaps even moreso than APS and AIP, \textbf{AAAS is an underutilized resource within the HEP community.} Further details are available on the AAAS website \cite{AAAS}.

It should be noted that the HEP community advocacy opportunities described so far, including those described in detail in Ref.~\cite{cef06paper1}, strongly revolve around the regular U.S. budget calendar. However, many non-funding legislative issues important to members of the HEP community are inherently irregular (\textit{i.e.} they do not follow funding timelines). In order to be serious about advocating for non-budgetary topics, \textbf{the HEP community must start to avail itself of advocacy pathways that do not rely on the federal budget calendar.}

\subsection{Opportunities for Expanding the Use of Existing Resources}

The activities of the broader physics community described above are all currently active, but throughout the proceedings of CEF06 it has become clear that awareness of the existence of these resources or the means to utilize them are poorly known of within the HEP community. Strengthening connections to these already active organizations and efforts would be beneficial to the HEP community by enabling community members to advocate for issues that impact them and the field overall. It is important to understand how the HEP community can interface more directly with these other organizations. A summary of specific items raised during conversations in this area during the course of Snowmass discussions is below.

\begin{itemize}
    \item Effort should be taken to increase knowledge of the available resources and other groups with these resources. Many people expressed a lack of awareness of the resources available to them. Talks at APS and APS DPF meetings could be a means to raise awareness. 
    \item A limitation of APS, AIP, and AAAS is that one nominally is required to be a dues-paying member to access all of their resources. However, these groups have been receptive to helping the HEP community in the past, so going through APS DPF or the users groups could be a path to gaining access to these resources for the whole community. 
    \item While the APS DPF executive committee does represent HEP broadly, there is no straightforward mechanism for community members to use this group as a route to providing feedback about APS advocacy. The creation of feedback mechanisms, to provide a connection between the HEP community and APS advocacy groups via APS DPF leadership, should be considered.
    \item Organizers of the HEP community advocacy effort have worked with APS government relations experts in the past, but there has not been a strong connection between these groups in recent years. It could be very beneficial to utilize APS DPF to build up that connection and to explore additional connections between HEP users groups and APS government relations experts. 
    \item Current government relations experts who work with the HEP community advocacy effort are experts in funding advocacy. Expanding the network of resources available to the HEP effort to include connections to APS, AIP, and AAAS would enable the community to access experts in additional areas.
    \item It would be beneficial to create a communication method for community members to express concerns about policy issues that affect them or the community at large. This raises difficult questions such as: who would be the owners of \textit{e.g.} a community-wide mailing list? And, who would act in what manner when concerns are raised? 
    \item It would be helpful to create an HEP-wide communication tool, such as an email list, that could be used to make community members aware of ongoing advocacy efforts. APS and APS DPF send out emails currently, but only members of those organizations receive them, and many HEP community members do not pay attention to them. Again this raises the question of who would maintain such a resource.
    \item  The users groups are active in funding advocacy, but they have been less involved in advocacy for other areas. DEI advocacy groups could be formed within the users groups. Such groups would benefit from strong connections to APS, AIP, and AAAS to help enable their resources to be more broadly available within the HEP community. 
\end{itemize}
\section{Summary of Discussions on HEP Community Advocacy}
\label{sec:discussion}

Advocating for areas other than HEP funding was widely discussed as part of the proceedings of CEF06. Assuming that there is consensus that such areas are worthy of our support, a key question that must be addressed moving forward is: Should we, as the HEP community, directly advocate for areas other than HEP funding, or should we exclusively rely on the advocacy efforts of external groups and strengthen our connections to those groups? We, the authors of this paper, do not offer an answer to this question, but instead summarize relevant considerations that we hope can be used for a wider community discussion on this topic.

We note that the framework for these discussions were originally motivated by two Snowmass ``Letters of Interest'' in this area submitted to CEF06:

\begin{enumerate}
    \item Culture change is necessary, and it requires strategic planning \cite{LOI_1}
    \item Non Funding Government outreach \cite{LOI_2}
\end{enumerate}

\subsection{Policy Areas Impacting the U.S. HEP Community}

The following specific areas were identified as potential topics of targeted, explicit, advocacy due to their relevance to the quality of life of community members and the collective science output of the field of HEP.

\begin{itemize}
    \item \textbf{DEI:} Issues of DEI are of the utmost importance and impact the whole field. The hardships of underrepresented minorities can not be understated, and the lack of diversity in our field directly affects the science output of the field.
    \item \textbf{Immigration policy:} The impact of policy on international scientists working in our field ranges from restrictions to the total exclusion of some international groups. Many international scientists living in the U.S. lack security in their legal status, meaning they have less stability in their career prospects compared to U.S. citizens. This is damaging to individuals in our community, as well as to the U.S. HEP community overall because it has the net effect of drawing talent away. Immigration issues are particularly impactful to early career members of the community because they are less likely to have permanent residence and generally have less support through their institutions in cases where public policy affects their ability to live and work in the U.S. This issue has an especially direct impact on the science output of the field.
    \item \textbf{STEM education:} Without strong support for STEM education, U.S. science would not retain or grow its current workforce and would not remain a leader in world science. This issue also has a direct impact on the science output of the field.
\end{itemize}

\subsection{Choice of Topics for Advocacy}

Current HEP community advocacy (for funding) is specifically nonpartisan and uses,  where possible, apolitical motivations for supporting HEP and basic research. The HEP community has a history of benefiting from bipartisan support. As noted above, as part of current advocacy efforts there is implicit advocacy for STEM, DEI, and international science -- these are topics for which there is also generally broad bipartisan support. However, the specifics of policy decisions and bills in these areas can be highly politicized and divisive. During the discussions of CEF06, concerns were raised on the potential impact to current HEP advocacy activities if the community were to also explicitly advocate for topics perceived to be partisan in nature. 

Any level of engagement in advocacy in these areas should follow from robust community consensus -- how this might be built is discussed below in Section \ref{sec:consensus}. Supposing that community consensus does exist that action should taken by the community in these areas, there would likely be insufficient resources to advocate for all topics that impact the HEP community. Therefore criteria would be required for determining the subset of topics for which the community would advocate. One possible criterion discussed is the immediacy of an issue's impact on the HEP community at large or members of the community. Identifying and quantifying impact in itself are nontrivial tasks, however, and some issues directly affect the community by their cumulative individual impact, which can be difficult to assess. 

An additional case was discussed, which is advocacy in response to an issue that rises to the level of a moral emergency. In this case, the issue at hand might not be directly impactful to the work of the HEP community, but rather directly impact the moral standards of the community. This hypothetical again presupposes the ability of the community to clearly establish consensus or lack thereof, whether on the impact of a particular issue or the moral standards which the community would act to defend.

\subsection{Defining Community Consensus}
\label{sec:consensus}

\noindent The point of community consensus, and how it should be defined, was a recurring theme through the CEF06 discussions in this area. Foremost among the questions raised was: \textbf{to what extent is consensus required for community members to engage in advocacy in any particular area?} We note that while, ideally, the community would have well-documented indications of consensus before any individuals speak on behalf of the community on any topics, in practice this is an unrealistic goal. Indeed, as noted above, existing advocacy efforts already include implicit advocacy for some non-funding topics and the specifics of the strategy implemented to advocate for funding are not subject to the approval, blessing, or review of the community at large. However, it is true, and critically important, that the broader themes of our funding-based advocacy are driven by the contents of the P5 report, which has strong, well-documented community support. It has been suggested that the construction of an analogous document could be used as the basis for general guidance on advocacy for non-funding topics, provided that it receives similarly strong, well-documented community support. We note that materials being produced within the Community Engagement Frontier (for example, within the topical group for Diversity, Equity, and Inclusion, CEF03 \cite{CEF03}) may provide a good groundwork to produce such a document. 

The above points are driven by the question of whether or not the community should continue or expand its advocacy for non-funding topics which are broadly understood to be directly beneficial to the field and are generally perceived to be nonpartisan. In the case of potential advocacy for issues which \textit{are} generally perceived to be partisan issues, the bar for consensus building is raised substantially. We assert the need for very clearly outlined and agreed-upon mechanisms to determine which individuals and representative bodies within the community are given the authority to define community opinion in these cases. Furthermore, we assert that this is a discussion which should be embraced before the explicit need arises. 

During the discussions of CEF06, the question of how specifically consensus might be built was largely set aside, though we note that a recurring idea was that there is presently a lack of a community-wide body that is able to directly engage in advocacy-adjacent matters. None of the users groups (UEC, SLUO, USLUA) represents the entire field, and APS DPF, which nominally does represent the entire field, is limited in its ability to engage in advocacy because it sits under the larger umbrella of APS. Such a body would require government relations expertise to advise community members on the impact of particular policies or advocacy that could be undertaken to influence policy. APS has such expertise at their disposal, but in order to utilize their resources we would have to work within the constraints of APS priorities, which don't necessarily align with HEP community priorities. We note additionally that the cost of APS membership has the potential to pose an exclusionary burden to community members.

\subsection{APS Resources for Advocacy}

One solution to community advocacy on non-funding topics that was identified is to increase the utilization of the existing infrastructure maintained by APS. It is certainly the case that APS has invested significant resources into understanding how to effectively engage in such advocacy, and has developed a framework to enable community members to opt into their efforts. We note that a distinct advantage to this approach is that the APS community is larger than the HEP community, and therefore can have a larger influence through advocacy. Conversely, the fact that the HEP community is only one of many under the APS umbrella has the effect to obscure the inner-workings of APS government relations and to dilute the relative impact of HEP community opinions on developing APS community opinions. And, as noted above, there may be cases in which the priorities of the HEP community are not aligned with the priorities of the broader APS community. In such cases it would be beneficial to maintain infrastructure internally within the HEP community to engage in HEP-community-specific advocacy. We additionally note that questions were raised regarding the efficacy of APS advocacy efforts -- we were unable to determine how to gauge the success of specific campaigns undertaken by APS to influence policy. Overall our working group was in consensus that steps should be taken to support and to increase awareness of APS advocacy campaigns within the HEP community.

\subsection{Utilization of Current Framework for Additional Advocacy}

\noindent The discussion about hypothetical community advocacy on non-funding topics was largely motivated by the success of existing HEP community advocacy efforts and the observation that the infrastructure to engage in that advocacy could reasonably be expanded or adapted into a new infrastructure to engage in additional advocacy. The idea of having the two efforts so closely linked (hypothetically) carries distinct advantages and disadvantages.

The existing HEP community advocacy effort, anchored by the annual ``DC trip'', is aimed at advocacy for Congressional appropriations and is specifically timed to maximally align with the cyclic nature of the Congressional budget process. The tactics of existing efforts have been finely honed with the federal budget cycle and process specifically in mind. If our end goal is to influence non-funding policy, then the most appropriate tactics will be different. Therefore, we generally have concluded that utilizing the ``DC trip'' itself would not be the most effective means to advocate for other topics. Another mechanism would need to be developed to advocate on different issues effectively, though such an alternate mechanism could still utilize tools that have made past HEP community advocacy successful.

\subsection{Impact to Current (Funding) Advocacy}

\noindent It is important to acknowledge that any hypothetical future advocacy by the HEP community on ``new'' topics has the potential to adversely impact existing advocacy efforts, which are focused explicitly on the funding of the field. During CEF06 discussions two classes of potential impact on existing advocacy efforts were identified: internal impacts and external impacts.

As discussed in detail in Ref. \cite{cef06paper1}, the existing HEP community advocacy efforts, while very successful, are not perfectly robust. There are well-founded concerns about the continued availability of resources to sustain this effort into the future before one adds the possibility that resources be additionally utilized for advocacy on other topics. During discussions on this topic, concerns were raised about dividing the limited time and labor resources currently provided by community members between both funding and non-funding advocacy. There is a risk that doing so would have the effect of splitting focus, thereby diluting the impact of each effort. It was also suggested that this division of attention could pose an existential thread to existing advocacy efforts because the existing organization, which has grown organically over the past 35+ years, may not be flexible enough to adapt to a drastically new purpose.

Supposing that sufficient resources were identified for the community to sustain such a dual-purpose advocacy effort, there is the additional concern of the external impact that additional advocacy could have on existing community advocacy for funding. Specifically, concerns have been raised that our community funding advocacy could lose support from Congressional offices due to hypothetical stances that the community might take on other (non-funding) issues. This was discussed specifically in the context of recent immigration policy changes in the U.S. which have impacted our field. In this case it was argued that engaging in advocacy on this topic could be seen as compromising political neutrality, that these kinds of advocacy are better done through APS and DPF, and that it is important that the HEP community be seen as nonpartisan.  This is a real risk that should be comprehensively understood in advance of any decision to engage in additional community advocacy. However, we note that community members have asserted that there are more important metrics by which to judge the success of our field than funding. Choosing not to take a stance on current affairs or to engage in advocacy on these topics is itself taking a stance, and this action too has real consequences on the field that should be comprehensively studied. Early career researchers in our community increasingly value the climate of the field and some number will choose to leave because the community is not moving fast enough or far enough towards addressing issues that they view to be important. It is important for us to recognize that there are community members that do not value funding for our field above all other priorities, in particular when that funding comes at the expense of prioritizing social aspects of our community and cultivating a culture that supports individuals in our community. 

\subsection{Outreach to the Community}

\noindent Past and ongoing HEP community advocacy efforts have failed to engage a significant fraction of the community. This is a notable feature in the context of funding advocacy (and is discussed in Ref.~\cite{cef06paper1}), and important to revisit in the context of hypothetical additional advocacy on other topics. Historically, both institutions and individuals within our community have provided mixed messages on the importance of engaging in advocacy (and outreach) activities. This generally manifests itself as a bias in career progression and the granting process against individuals that are perceived to spend excessive time engaging in such activities. Simultaneously, the end result of community advocacy for funding of the field is widely acknowledged to be critical to its current and future successes. It is likely that the perceived disadvantages to involvement in advocacy for non-funding topics would not have an equivalent offset and thus would have a net negative effect on the career progression of individuals that choose to engage in these activities. We note that there seems to be a (slow) cultural shift towards acknowledging the value of individuals engaging in these community-oriented activities, but this has yet to be reflected in the formal institutions of our field. Additionally, in the case of advocacy, there are legal limits to extent to which the prioritization of these activities could be reflected in our institutions. Some employees (\textit{e.g.} those at national laboratories or community members serving on HEPAP) can be prohibited from engaging in policy advocacy, and the Hatch Act prohibits any federal funding (including grant funding and national laboratory employee salary) to support advocacy activities.

Finally, we note that there is a tendency for work in community engagement areas to fall on the individuals that are most impacted by the problems they are trying to solve. It is our assertion that the people most impacted by DEI and discriminatory immigration policy should not have to bear the responsibility of advocating for these policies to change. They impact the whole community and the whole community needs to participate in the choice of how to address them. And if any advocacy is a part of the plan to address them, then the whole community needs to participate in that advocacy. 

\section{Conclusion}
\label{sec:conclusion}

This paper summarizes the legislative topics that are impactful to the High Energy Physics community and what government engagement activities and resources exist to engage in advocacy on these topics. We have discussed how the community could use the currently underutilized resources from APS, AIP and AAAS more, and have presented a summary of the discussions of CEF06 on these topics. This includes a breakdown of the potential advantages and complications of setting up HEP community advocacy in this area. 

%%%%%%%%%%%%%%%%%%
%%% References %%%
%%%%%%%%%%%%%%%%%%
\bibliographystyle{unsrtnat}
\bibliography{Bibliography/common,Bibliography/main}

\end{document}